\documentstyle[prb,aps,epsf,multicol]{revtex}

\newcommand{\metmat}{{\bf G}}

\begin{document}

\title{Simulation of Material Properties Below the Debye Temperature:
  A Path-Integral Molecular Dynamics Case Study of Quartz}

\author{Martin H. M\"user}

\address{Institut f\"ur Physik, WA 331; Johannes Guntenberg-Universit\"at\\              55099 Mainz; Germany}

\date{\today}
\maketitle

\begin{abstract}
Classical and path integral molecular dynamics (PIMD) simulations
are used to study $\alpha$ and $\beta$ quartz in a large range of
temperatures at zero external stress.
PIMD account for quantum fluctuations of atomic vibrations, which
can modify material properties at temperatures below the
Debye temperature. 
The difference between classical and quantum mechanical
results for bond lengths, bond angles, elastic modulii, and some dynamical
properties is calculated and comparison to experimental data is done.
Only quantum mechanical simulations are able to reflect the correct
thermomechanical properties below room temperature.
It is discussed in how far classical and PIMD simulations can be helpful in
constructing improved potential energy surfaces for silica.
\end{abstract}

\begin{multicols}{2}

\section{Introduction}
Quartz is one of the most abundant and best studied minerals
on Earth; the structure and many other physical properties
are well understood~\cite{heaney94}. 
Classical molecular dynamics (MD) simulations have been particularly
successful in connecting atomic interactions between silicon and
oxgen atoms in the condensed phase
with structural and elastic properties of 
quartz~\cite{tsuneyuki88,beest90,tse91,somayazulu93,muser00}
and other silica polymorphs at finite 
temperatures~\cite{tsuneyuki88,tse91,tsuneyuki90,tse92}.
At zero temperature, this connection has also been done
by mere first-principle
studies~\cite{liu94,boer95,swainson95} or ab-inito methods incorporating
bulk-system information~\cite{beest90}.

Some of the numerical approaches reach a nearly perfect agreement with
experiment, especially for structural properties.
None of the calculations, however, include quantum effects of the ionic
motion. Quantum effects lead to an equilibrium structure
that is different from the ``classical'' equilibrium structure
as soon as the interactions between the atoms constituting a crystal
are not purely harmonic.
Path integral simulations are a convenient tool to compute such
quantum mechanical effects~\cite{muser95,martonak98,herrero00}, e.g., the
zero-temperature
lattice constant $a$(Ne$^{\rm cla})$ of classical Neon is 0.18$~\AA$ smaller
than the one of Ne$^{22}$, which again is about 0.19\% smaller than
$a$(Ne$^{20})$.~\cite{muser95}
Treating atomic motion quantum mechanically even effects 
covalent bonds in systems such as 
crystalline polyethylene~\cite{martonak98}
and silicon~\cite{herrero00}.

Quantum effects are certainly
less strong in silica than in rare gas solids such as in neon.
Thermal expansion can nevertheless be observed for $\alpha$-quartz
below the Debye temperature and
quantum effects have to be taken into account properly,
if we want to relate simulations and experiments in a meaningful way.
Correct quantum simulations will
converge to a zero expansion coefficient as absolute zero is approached,
while classical MD simulations will (nearly) always result in a
finite expansion coefficient even at $T = 0$~K thus violating the
third law of thermodynamics.
It is obvious
that structure and elastic constants are reflected well by
the model potential if these properties have been used to
fit the free parameters of the
potential energy surface - as done in the case of
the so-called BKS potential~\cite{beest90}.
It is therefore a much harder test for a model potential surface to yield
the correct thermal expansion at low temperature, because the anharmonic
interactions must be reflected accurately by the model potential. 
Overall, ab-initio calculations, MD simulations, and experimental data have
partially achieved such good agreement that estimating
quantum effects will play an important role in determining
the real merits of a potential energy surface.

In this study, path-integral molecular dynamics (PIMD) are used to determine 
the quantum effects on physical properties of quartz. Due to the
large computational demand of path integral simulations,
we confine ourselves to the use of only one potential energy surface,
namely the BKS potential~\cite{beest90}. It
has been particularly successful in reproducing silica properties
not only of $\alpha$-quartz~\cite{beest90,tse91}, 
but also of other silica polymorphs~\cite{tse91,tse92}
and of the glassy state~\cite{vollmayr96,badro98,horbach99}.
It is therefore rather plausible that the BKS potential predicts
accurately the shifts from classical results to quantum mechanical results.

Recently, $\beta$-quartz and $\beta$-cristobalite have been
investigated by means of path integral Monte Carlo (PIMC) 
simulations~\cite{rickwardt}.
The PIMC study, however, was only done for phases that are stable
at temperatures above $800$~K. Quantum effects are relatively small
at such high temperatures. Moreover, the resolution of our PIMD simulations
exceeds that of the emploied PIMC algorithm by orders of magnitudes.
E.g., the PIMD approach makes it possible to calculate
the ground-state equilibrium lattice constants with a resolution of more
than $0.001\,\AA$ for a given potential energy surface,
while the PIMC simulations had an uncertainty of typically $0.1\,\AA$.
This improvement in resolution is possible despite a strong
reduction in CPU time.
Furthermore, arbitrary parallelepiped simulation cells are permitted
in this PIMD study allowing to calculate all elastic constants.
The PIMC studies were confined to orthorhombic geometries and did not
allow calculations of any elastic constants due to large statistical error
bars. 

The remainder of this paper is organized as follows:
In Sec.~\ref{sec:method}, the PIMD method used in this study is described
along with some specific, technical details. In Sec.~\ref{sec:results},
results are presented for structural data, elastic constants, and dynamic
properties. Quantum mechanical results are compared to classical simulations
and experiment. Conclusions are drawn in Sec.~\ref{sec:conclusions}.

\section{Method}
\label{sec:method}

\subsection{Path Integral Molecular Dynamics in the Constant Stress Ensemble}

Although path integral Monte Carlo (PIMC) is usually used to estimate
quantum effects in solids~\cite{chakravarty97,nielaba97,marx99},
path integral molecular dynamics (PIMD)~\cite{tuckerman93} arise
as a more natural choice for long-range Coulomb interactions. 
The Ewald sum~\cite{frenkel96}
has to be evaluated only once per
time step in PIMD as opposed to each single local move in PIMC.
Moreover, global moves in which the shape and size of the simulation cell
are varied, are at basically no extra cost in molecular dynamics
simulations, while Monte Carlo requires the evaluation of the net energy
for each single trial move of the strain tensor.

In the path integral formulation of quantum
statistical mechanics~\cite{feynman65},
a quantum point particle at temperature $T$ is
represented by a closed polymer at temperature $P\,T$, in which adjacent
beads interact via harmonic springs. The stiffness of the springs increases
with decreasing thermal de Broglie wavelength $\lambda$
that a free particle would
have at temperature $P\,T$. Studying this model in a molecular dynamics
simulation, would require small time steps
in the quantum limit $P \to \infty$,
if the dynamical masses of the polymer beads were chosen to be identical with
the physical mass $m$ of the quantum particle. However, it is possible to
adjust all intra-molecular vibrations to similar time scales
if the equations of motion are expressed in a convenient representation
and appropriate ``dynamical'' masses are attributed to the 
beads.~\cite{tuckerman93} In this study, the coordinates are represented
in terms
of eigenmodes of the free particle. The dynamical masses 
$\tilde{m}_\omega$ attributed
to the motion of the eigenmode $\omega$ are usually chosen according to 
$\tilde{m}_\omega/m = k_{\rm E} / (k_{\rm E} + k_\omega)$,
with $k_{\rm E}$ the coupling of an atom to its lattice site
in the Einstein model of solids and $k_\omega$ the spring constant associated
with eigenmode $\omega$. This choice of dynamical masses allows for efficient
sampling of all degrees of freedom, because
all modes move essentially on the same time scale.
The dynamical mass of the center-of-mass motion of the
polymer $\tilde{m}_0$ is of course identical with the real mass $m$.

The motion of the simulation cell is constrained to symmetric
strain tensors, but otherwise done as is in the classical Parrinello-Rahman 
method~\cite{parrinello80}. 
The dynamical mass $W$ associated with the motion
of the simulation cell geometry is again chosen such that a typical
oscillation time of the box is close to a typical oscillation time of
a silicon or oxygen atom. The choice of $W$ merely controls the efficiency
of the sampling but leaves meaningful observables uneffected.

One advantage of the Parrinello-Rahman method is the possibility to
determine  all elastic constants at zero external stress.\cite{parrinello82}
This is done by using appropriate relations between 
strain fluctuations and mechanical compliances. 
It is important to note that only {\it isothermal}
strain fluctuations are accessible in PIMD simulations.
A constant enthalpy simulation of the isomorphic classical representation
would not translate into conserved enthalpy of the quantum crystal.

In principle, our method is closely related to a recently proposed
PIMD scheme for constant-strain constant-temperature 
simulations~\cite{tuckerman98,martyna99}. The special representation used 
here as well as omitting the thermostat included in the equations of
motions in Ref.~\cite{tuckerman98,martyna99}, anticipates to briefly
review the final result.
In order to do this, we represent the coordinate 
${\bf R}_{i t}$ of particle $i$ at imaginary time $t$ 
as a product of a scaled coordinate ${\bf r}_{i t}$
and the time-dependent
(symmetrical) matrix $h$, which contains the shape and
the volume $V = \det h$ of the simulation cell:
\begin{equation}
R_{i t \alpha} = h_{\alpha\beta} r_{i t \beta}.
\end{equation}
The values of $r_{i t \alpha}$ are constrained to values
$ 0 \le r_{i t \alpha} < 1$.
The components of the metric tensor $G$ are defined as
$G_{\alpha\beta} = h_{\alpha\mu} h_{\mu\beta}$ where 
summation convention over Greek indices enumerating spatial dimensions
is implicitly assumed.
It is then convenient
to  express the equations of motion for the scaled coordinates
in reciprocal Fourier space, namely in terms of
coordinates
\begin{equation}
\label{eq:fourier}
\tilde{\bf r}_{i\omega} = {1\over \sqrt{P}}
\sum_{t=1}^P {\bf r}_{it} \exp\left({2\pi{\bf i}\over P}\omega t\right)
\end{equation}
for which the motion of the free particles is diagonalized.
Introducing $k_{i\omega} = 4m_i \sin^2(\pi\omega/P)/(\beta\hbar/P)^2$,
allows to represent
the equations of motion in a rather condensed form:
\begin{eqnarray}
\label{eq:eqa_mot_1}
m_{i\omega} \ddot{\tilde{r}}_{i \omega \mu} & = &
\tilde{m}_{i\omega} (\metmat^{-1})_{\mu\nu} \dot{G}_{\nu\sigma}
\dot{\tilde{r}}_{i \omega \sigma} - k_{i\omega} 
\tilde{r}_{i \omega \mu} + \nonumber \\
& & {1 \over \sqrt{P}}
\sum_{t} e^{{2\pi {\bf i}\over P} \omega t}
\sum_{j \ne i} 
{\partial v_{ij} \over \partial R_{(ij)t}}
{r_{i t \mu} - r_{j t \mu} \over R_{(ij)t}}
\\
\label{eq:eqa_mot_2}
PW \ddot{h}_{\mu\nu} & = &
\sum_{i\omega} \tilde{m}_{i\omega} \dot{\tilde{r}}_{i\omega\nu} 
\dot{\tilde{R}}_{i\omega\mu} -
\nonumber \\ & &
\sum_{it}
{m_iP^2\over \beta^2\hbar^2}
(r_{i t \nu}-r_{i\, t-1\, \nu})(R_{i t \mu}-R_{i\, t-1\, \mu}) +
\nonumber\\ & &
\sum_{it} \sum_{j>i}
{\partial v_{ij} \over \partial R_{(ij)t}}
R_{(ij)t\mu} R_{(ij)t\nu}
\end{eqnarray}
with $v_{ij}$  a two-particle interaction potential between particle
$i$ and $j$ and ${\bf R}_{(ij)t}$ the vector connecting
particle $i$ and $j$ at imaginary time $t$
emploing minimum image conventions.

Despite the well-known disadvantages of 
Langevin-type thermostats~\cite{frenkel96,schneider78},
correlation times turned out to be particulary small when all
degrees of freedom (including the geometry of the simulation cell)
were weakly coupled to a friction force linear in velocity and to
a corresponding random force.~\cite{schneider78}
Chosing the damping term $\gamma$ of the
Langevin dynamics to be $\gamma = 0.01 \,dt$ and $dt = 0.04 t_{\rm char}$
with $t_{\rm char}$ the (smallest) characteristic time-scale of the system,
systematic errors were made much smaller than statistical errors.
Moreover, the ergodicity problems, which is
inherent to some PIMD algorithms~\cite{tuckerman93},
can be most easily overcome with a Langevin thermostat.
The average ``dynamic'' kinetic energy $\langle T_{\rm dyn}\rangle$
and the fluctuation of $T_{\rm dyn}$ in the
PIMD approach described above, are utterly sensitive
to bad choices of $\gamma$ and $dt$. In the correct limit, one obtains
$\langle T_{\rm dyn}\rangle/N = d k_{\rm B}TP/2$ per degree
of freedom and the associated specific heat (fluctuation)
$\langle \delta T^2_{\rm dyn}\rangle/Nk_{\rm b}T^2P^2 = d k_{\rm B}/2$
with $d$ the spatial dimension of the system.
This sensitivity can be
used to determine the accuracy of the simulation resulting
in reliable thermodynamic expectation values of 
other observables. Note that only independent components of the
Fourier transform $\tilde{{\bf r}}_{i\omega}$ need to be 
thermostated and considered for the calculation of the dynamic kinetic
energy.

\subsection{Dynamical information}

Imaginary-time path-integral methods do not allow direct calculation
of dynamical  properties. 
While the complete dynamical
information is obtained in imaginary-time correlation functions 
in principle~\cite{baym61}, the inverse Laplace transform that one needs
to carry out in order to assess real-time correlation functions is
numerically unstable. 
A generalization of the PIMD method, the centroid
molecular dynamics (CMD) method~\cite{cao93}, appears to give a more
direct link between exact quantum dynamics and CMD. The basic idea of
CMD is to propagate the center of mass of a PIMD quantum chain such that
the internal degrees of freedom of the quantum chain are in their 
thermodynamical equilibrium. This makes the center of mass move on
an effective classical potential surface. Time correlation functions
can then be formulated in terms of centroid coordinates and averaged
during the simulation. The Fourier transform
of the centroid correlation function $I_{\rm c}(\omega)$ and
the real quantum mechanical spectral function $I(\omega)$ are 
related linearly by
\begin{equation}
\label{eq:centroid_spec}
I(\omega) = 
n(\beta\hbar\omega) I_{\rm c}(\omega)
\end{equation}
with
\begin{equation}
n(\beta\hbar\omega) = {\hbar\beta\omega\over 2} 
\left(1 + \coth{\hbar\beta\omega\over 2} \right).
\label{eq:norm}
\end{equation}
Eq.~(\ref{eq:centroid_spec}) is exact for harmonic systems as well as
in the classical
limit. Recently, much progress has been made in 
formulating  rigorous relations between the dynamics of
path-integral centroid variables and true dynamics as well as
systematic correction to the CMD method sketched above.~\cite{jang99} 

\subsection{Model Specific Information}

In the present simulation, $v_{ij}$ in 
Eqs.~(\ref{eq:eqa_mot_1},\ref{eq:eqa_mot_2})  corresponds to
the BKS interaction potential~\cite{beest90}
between particles $i$ and $j$ including  the
Coulomb energy among other contributions.
For the evaluation of the Ewald sum in arbitrarily shaped 
parallelepiped simulation cells, a recently proposed 
algorithm~\cite{wheeler97} was used and constrained to
symmetric matrices $h$ resulting in a 30\% reduction
of CPU time to evaluate the Ewald sum.
The non-Coulombic interactions were cut off at a distance 
$r_{\rm c} = 9.5$~$\AA$.

In the following, it will be distinguished between classical and
quantum mechanical results. Classical results are obtained
by simply chosing $P=1$ in a PIMD simulation. Exact quantum
results require taking the limit $P\to\infty$ in principle.
It is well known that the so-called primitive decomposition
of the density matrix, upon which the PIMD algorithm is based,
invokes systematic errors in the partition function and observables
that vanish proportionally to $1/(TP)^2$~\cite{muser95}.
It is therefore important to
chose $P$ large enough so that quantum effects are well reflected.
On the other hand, $P$ should not be too large, which would result
in large statistical error bars.

\begin{figure}[hbtp]
\begin{center}
\leavevmode
\hbox{ \epsfxsize=70mm \epsfbox{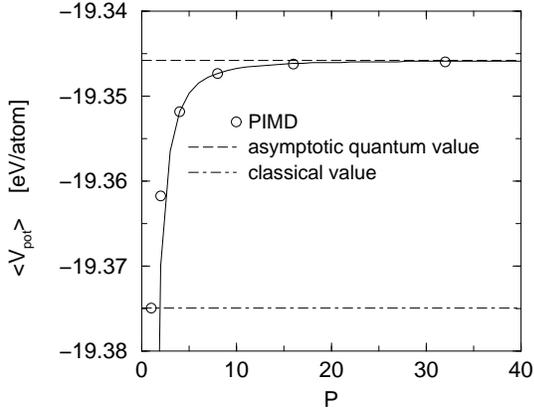} }
\begin{minipage}{8.5cm}
\caption{Average potential energy $\langle V_{\rm pot} \rangle$
for different Trotter numbers, $P = 2^n$ with $n = 0,\cdots,5$, at
$T = 300$~K. Straight line is a fit for large $P$ including corrections
to the asymptotic quantum value
in the order of $1/P^2$. Error bars smaller than 1~k$_{\rm B}$K.
\label{fig:vpot_room}
}
\end{minipage}
\end{center}
\end{figure}

In order to assess at what values of $TP$ one can expect convergence
to the quantum limit, the average potential energy 
$\langle V_{\rm pot} \rangle$ is calculated for various values of $P$
at the temperature $T = 300$~K.
The results are shown in Fig.~\ref{fig:vpot_room}. 
For $PT > 1200$~K, extrapolation to the quantum limit is possible 
with correction in the order $1/(TP)^2$. Combinations satisfying
$PT > 4,800$~K basically correspond to the quantum limit.

Depending on the property of interest, it is necessary to extrapolate
to the quantum limit with a $1/P^2$ corrections. This concerns
particulary structural data, such as lattice constants, which can be obtained
with high resolution. Elastic constants and dynamical information, however,
are plagued with relatively large statistical error bars.
For these observables,
statistical uncertainties are much larger than
systematic deviations and results obtained with $PT > 2,400$~K are
referred to as quantum limit.

The total number of atoms used in the simulations was $N = 1080$.
The linear box dimensions typically are $24.9\,\AA$, $25.9\,\AA$, and
$21.9\,\AA$ along the $a,b$, and $c$ axis, respectively.
In order to obtain elastic constants with an accuracy of about $3$~GPa,
60,000 MD steps of length 1~fs have to be performed, which takes about
one day on an Intel II processor. For quantum simulations, the numerical
effort has to be multiplied with Trotter number $P$.
\section{Results}
\label{sec:results}

\subsection{Structural Properties}
\begin{figure}[hbtp]
\begin{center}
\leavevmode
\hbox{ \epsfxsize=70mm \epsfbox{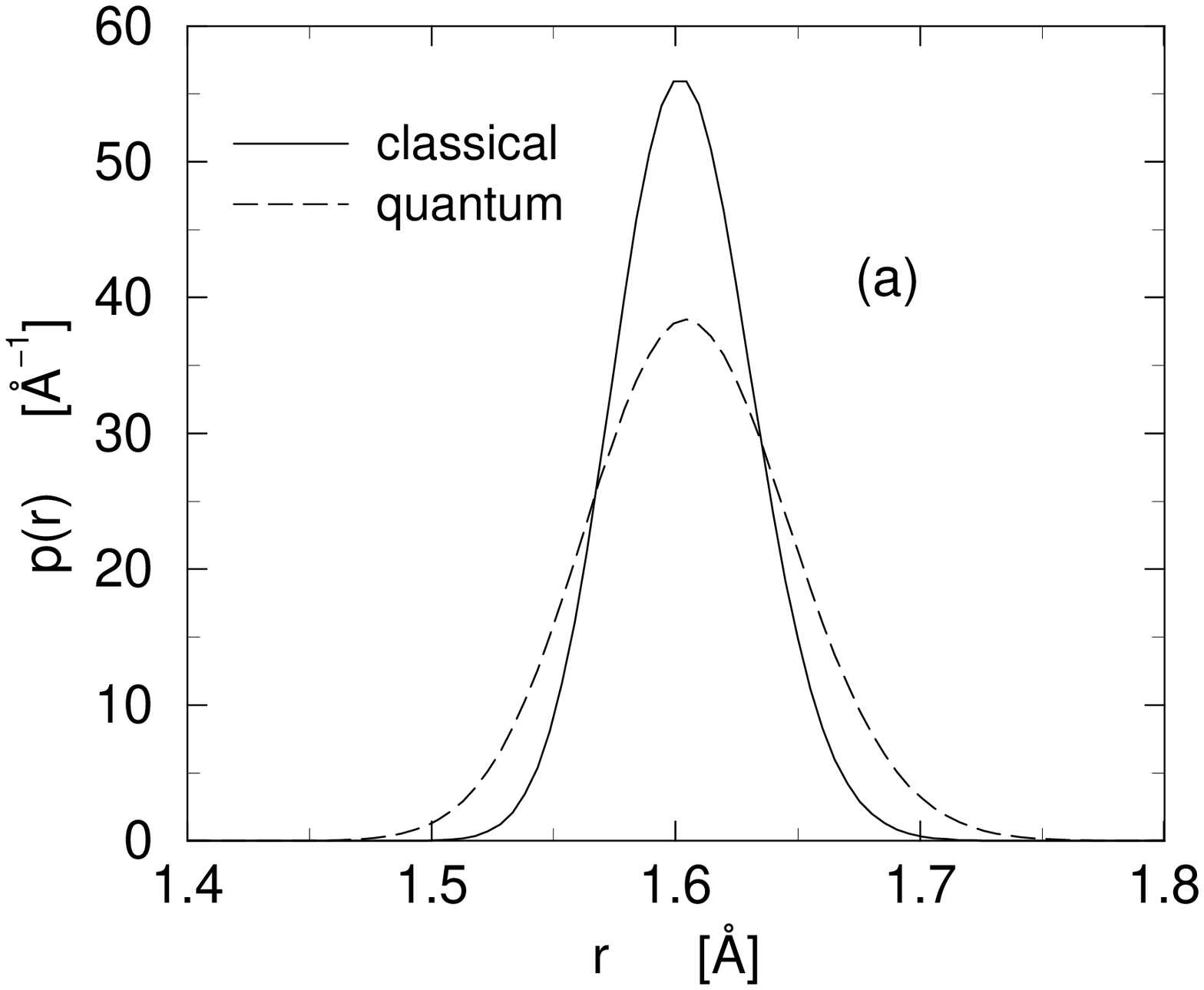} }

\hbox{ \epsfxsize=70mm \epsfbox{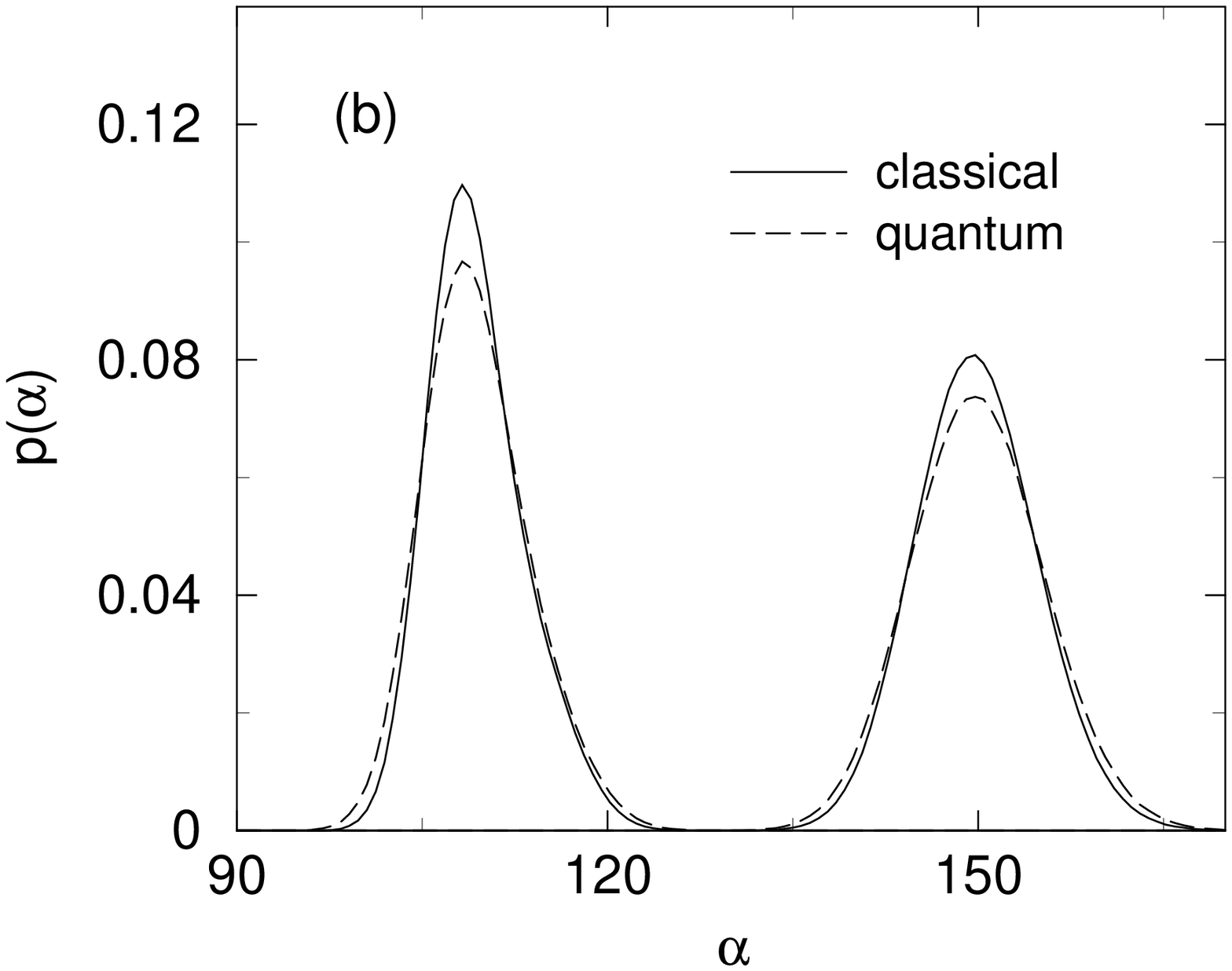}
} 
\begin{minipage}{8.5cm}
\caption{
a) Probability density $p(r)$ to find an oxygen atom in a distance $r$ from a
   silicon atom.
b) O-Si-O (left) and Si-O-Si (right) bond angle distribution  function 
   $p(\alpha)$. Solid lines reflect classical simulations, dashed lines
   represent quantum mechanical simulations. Temperature $T = 300$~K.
\label{fig:bond_room}
}
\end{minipage}
\end{center}
\end{figure}

Two temperature regimes are particularly interesting to study,
namely room temperature and temperatures near absolute zero.
We first start with a discussion of quantum effects at room 
temperature, $T = 300$~K.
This temperature is already well below the Debye temperature $T_{\rm D}$
of quartz. $T_{\rm D}$ of $\alpha$-quartz as determined by specific heat
measurements is a strongly temperature-dependent function~\cite{striefler75}:
At $T = 0$, $T_{\rm D} \approx 550$~K, while at room temperature
$T_{\rm D} \approx 1,000$~K.
Some quantum effects can therefore be 
expected to be relatively strong at room temperature.
In Fig.~\ref{fig:bond_room}, one can see that the main quantum effect
at room temperature is
the quantum mechanical freezing of the Si-O bond.
This freezing is reflected by the fact that the
distribution $p_{\rm SiO}(r)$ of the Si-O bond is much broader for the 
quantum mechanical study than for the classical study 
(Fig.~\ref{fig:bond_room}a). In fact, the quantum mechanical
$p_{\rm SiO}(r)$ barely
changes its form when the temperature is lowered further.
This is an indication that Si-O bonds are in 
their quantum mechanical ground state.
On the other hand,
Si-O-Si bond angles as well as O-Si-O bond angles
do not differ considerably between classical and
PIMD simulations (Fig.~\ref{fig:bond_room}b). 

\begin{figure}[hbtp]
\begin{center}
\leavevmode
\hbox{ \epsfxsize=70mm \epsfbox{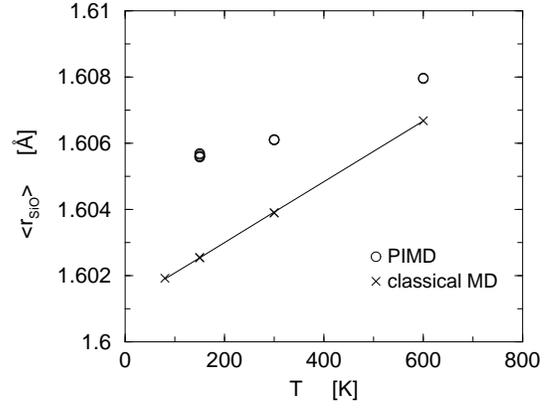} }
\begin{minipage}{8.5cm}
\caption{Average Si-O bond length $\langle r_{\rm SiO} \rangle$
as a function of temperature. A linear fit of the classical data
indicates an average equilibrium distance of 1.6017~$\AA$.
\label{fig:bond}
}
\end{minipage}
\end{center}
\end{figure}

\begin{figure}[hbtp]
\begin{center}
\leavevmode
\hbox{ \epsfxsize=70mm \epsfbox{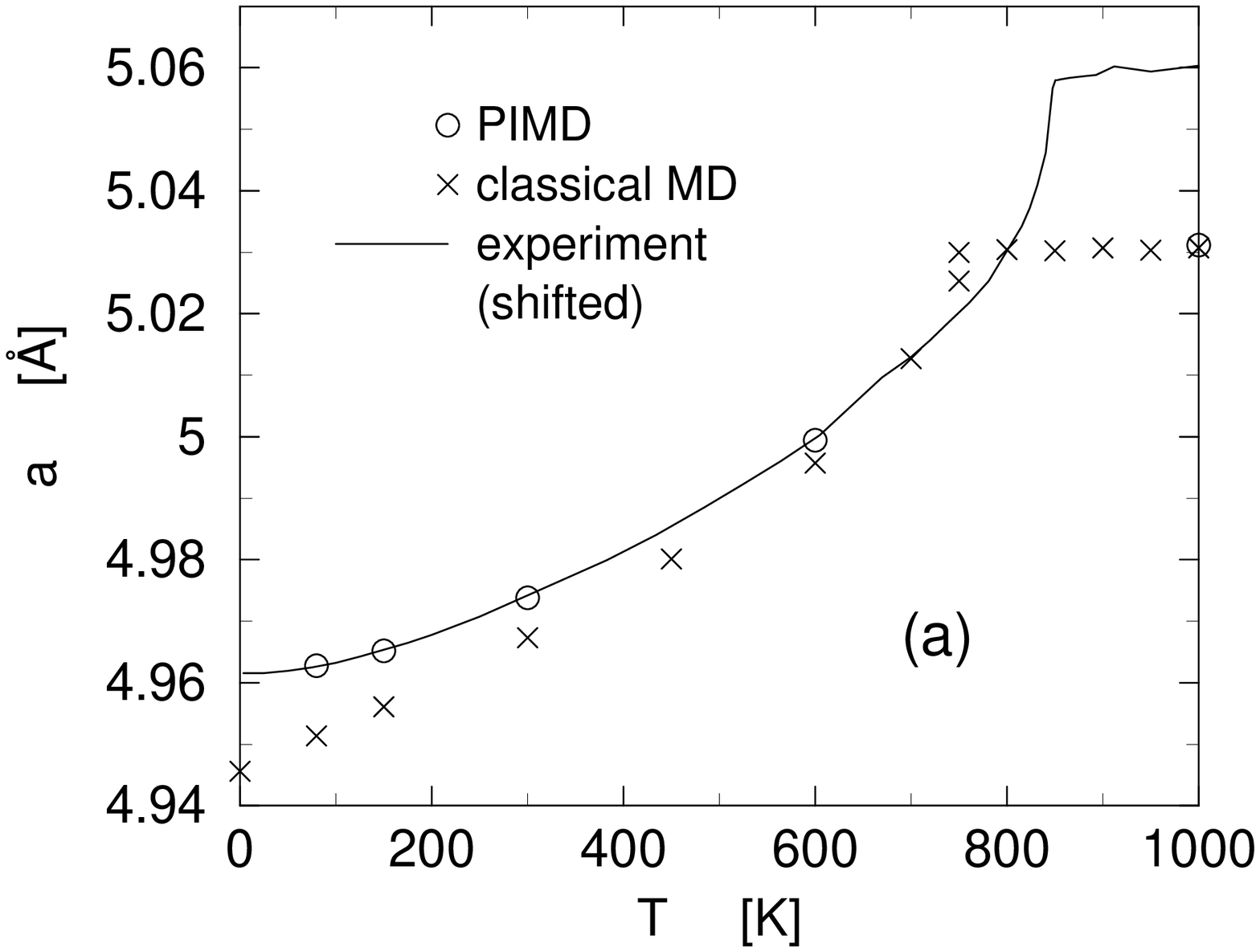}}

\hbox{ \epsfxsize=70mm \epsfbox{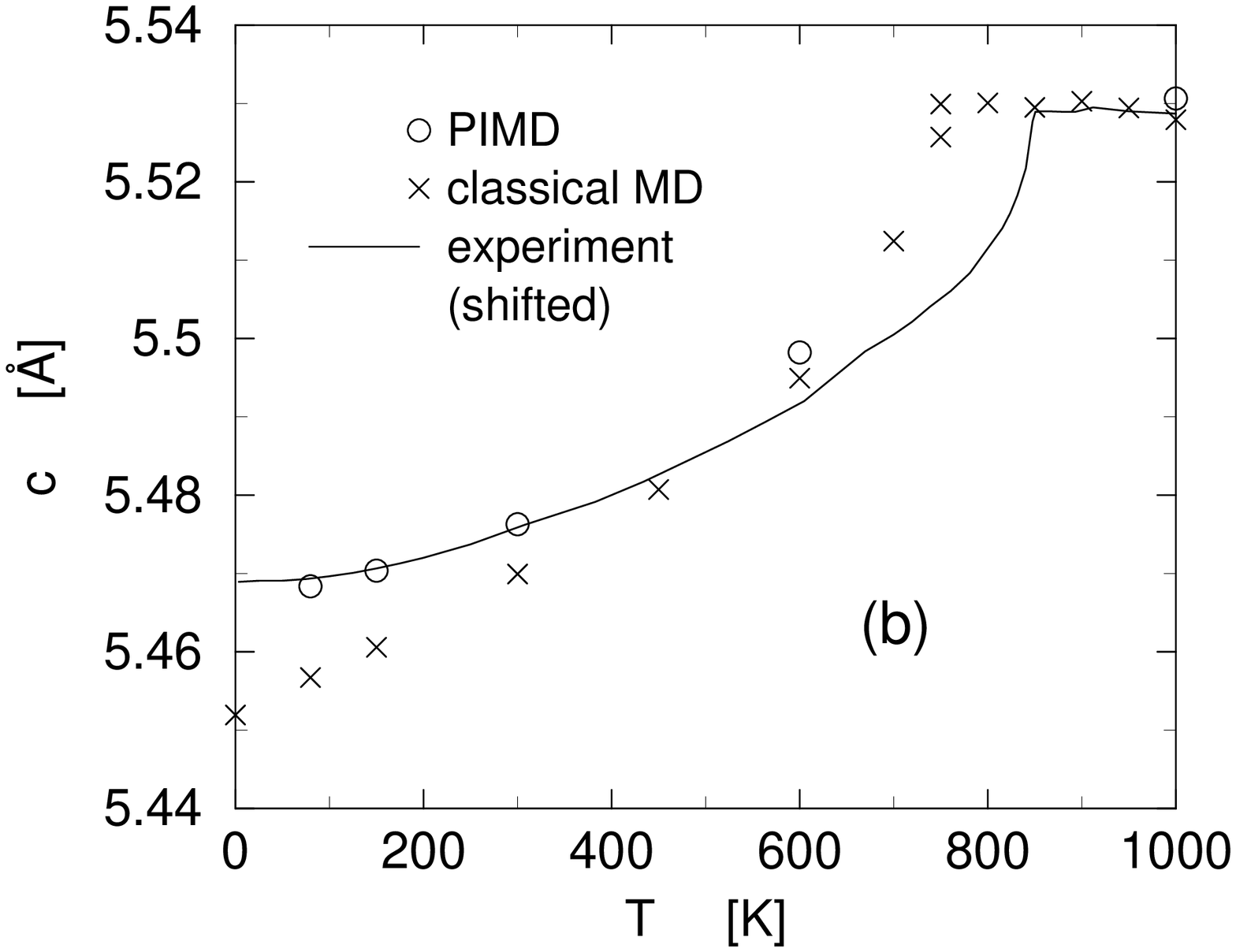}}
\begin{minipage}{8.5cm}
\caption{
Lattice constants of $\alpha$-quartz at ambiant pressure as a function
of temperature.
a) Experimental values for the $a$-axis are shifted to
larger values by $0.06\,\AA$.
b) Experimental values for the $c$-axis are shifted to larger values
by $0.07 \,\AA$.
Error bars in all cases smaller than 100~fm.
\label{fig:latt_const}
}
\end{minipage}
\end{center}
\end{figure}

Fig.~\ref{fig:bond} and
Fig.~\ref{fig:latt_const} show the effect of the quantum mechanical ionic
motion on simple structural properties. The average Si-O bond length
$r_{\rm SiO}$ is shown in Fig.~\ref{fig:bond} as a function of temperature,
while the lattice constants $a$ and $c$ are shown in 
Fig.~\ref{fig:latt_const}. In all cases, it is noticeable that the
quantum mechanical values are larger than the classical equilibrium lengths.
The effects are relatively small, but clearly within the resolution of the
simulations. While $r_{\rm SiO}$ only differ by 0.19\% at $T = 150$~K,
the lattice constants differ by 0.35\% in the case of both the $a$ axis 
and $c$ axis of $\alpha$-quartz. This means that the
``excess'' quantum volume can be attributed to both the SiO bond length
and quantum fluctuations of the so-called rigid unit 
modes~\cite{dove97,welche98}.

In the case of the lattice constants, Fig.~\ref{fig:latt_const},
direct comparison can be made to experimental data~\cite{carpenter98}.
The difference in the lattice constants 
between quantum mechanical calculations and
experiment is about $0.06\,\AA$
for the $a$ axis and $0.07\,\AA$ for the $c$ axis.
This difference is much larger than the discrepancy between 
classical and quantum mechanical simulations. It is, however, obvious
that only quantum mechanical simulations reflect qualitatively
the right low-temperature behaviour: While classical simulations
result in a finite expansion coefficient even at absolute zero,
the PIMD simulations lead to a vanishing expansion coefficient.

There is, however, a very good agreement in the low-temperature
thermal expansion along the $c$ axis between PIMD simulation and experiment
up to about 600~K, which shows that even anharmonic effects are
well reflected by the BKS potential as long as the system is still far
away from the $\alpha$-$\beta$ transition. Near the transition, however,
the agreement becomes siginficantly less good.
The ``jump'' in $c$ at the transition
is absent in the simulation. Thus, an important ingredient in the
potential energy surface is missing. Note that classical simulations
based on both the BKs and the TTAM potential
do not reflect the anomaly in the
$c/a$ ratio, which has been observed experimentally~\cite{muser00}.
Thus, while harmonic and low-temperature anharmonic effects are well
described by the BKS potential,
there seems to be the need to reflect many-body effects as well.

\begin{figure}[hbtp]
\begin{center}
\leavevmode
\begin{minipage}{8.5cm}
\hbox{ \epsfxsize=70mm \epsfbox{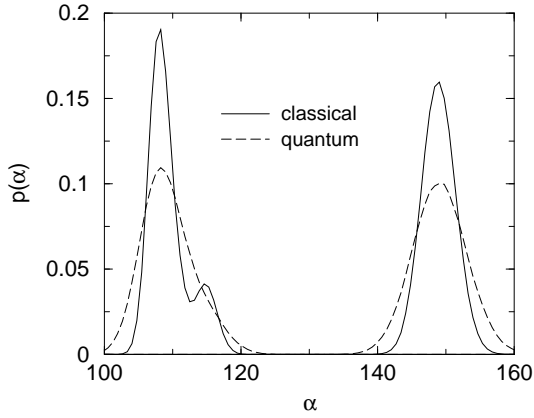} }
\caption{O-Si-O (left) and Si-O-Si (right) bond angle distribution
at temperature $T = 80$~K.
\label{fig:angle_80}
}
\end{minipage}
\end{center}
\end{figure}

\begin{figure}[hbtp]
\begin{center}
\leavevmode
\hbox{ \epsfxsize=70mm \epsfbox{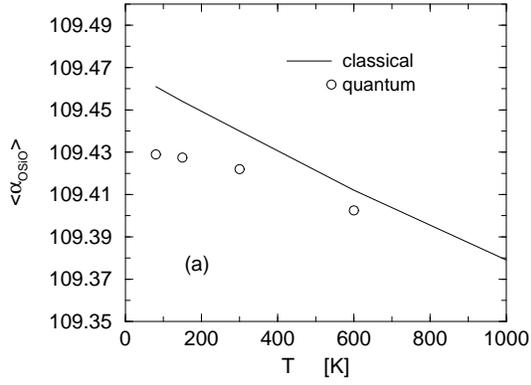} }
\hbox{ \epsfxsize=70mm \epsfbox{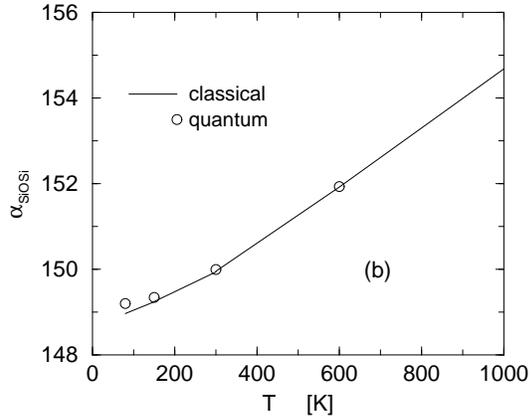} } 
\begin{minipage}{8.5cm}
\caption{
Mean bond angle as a function of temperature for classical (straight line)
and quantum mechanical (circles) simulations.
a) Si-O-Si bond.
b) O-Si-O bond.
\label{fig:angle_temp}
}
\end{minipage}
\end{center}
\end{figure}

Fig.~\ref{fig:angle_80} and Fig.~\ref{fig:angle_temp}
give detailed information on the bond angle distribution and their
quantum effects. In Fig.~\ref{fig:angle_80}, the Si-O-Si  and O-Si-O
angle distributions are shown exemplarily at a temperature $T = 80$~K.
Fig.~\ref{fig:angle_80} confirms the picture that the local structure
in $\alpha$-quartz does not correspond to tetrahedra. This can be concluded
from the existence of two peaks in the classical O-Si-O bond angle distribution 
and a broadened shoulder in the quantum mechanical O-Si-O bond angle
distribution. The difference in quantum mechanical and classical
mean bond angles is rather small, yet, noticeable, e.g., the classical
average O-Si-O  bond angles approaches the ideal tetrahedra angle of
$109.471^o$ much closer than the quantum mechanical simulation.
The difference of about $0.04^o$ between the two approaches can 
be clearly resolved. The effect is much larger for the Si-O-Si
bond angle, namely $0.2^o$ at $T = 80$~K, but less obvious
(Fig.~\ref{fig:angle_temp}b)
because of the strong temperature dependence of 
$\langle\alpha_{\rm SiOSi}\rangle$.

\subsection{Elastic properties}
\label{subsec:ela_const}
Just like other properties can elastic constants be expected to differ
between classical and quantum mechanical treatments. In order to
calculate classical constants at zero temperature, it is sufficient to
calculate the second derivative of the ground state (potential) energy
with respect to the stress tensor, resulting in the so-called Born
expression for elastic constants:~\cite{born54}
\begin{equation}
C_{\alpha\beta} =\partial^2 \langle V(T,\epsilon) \rangle /
 \partial \epsilon_\alpha \partial \epsilon_\beta,
\label{eq:born}
\end{equation}
where the derivative is evaluated at zero strain.
At finite temperatures, it is not sufficient to generalize this expression
by simpling taking the thermal expectation value of the right hand side.
It has been pointed out correctly~\cite{squire69} that
the free energy surface ${\cal F}(T,\epsilon)$ should be considered instead
of $V(T,\epsilon)$. This generalization leads to different estimators
of the elastic constants when evaluated in the (NVT) ensemble.
The main effect of this generalization is that fluctuations of the stress
tensor need to be considered on top of the Born term described in
Eq.~\ref{eq:born}. These fluctuations usually lead to a reduction
of the elastic constants.
Unlike classical fluctuation terms, quantum mechanically calculated
terms will not vanish as the temperature approaches absolute zero.
Among other effects,
this will lead to different elastic constants for quantum mechanical
and classical systems~\cite{schoffel01}.
In the case of silicates, however, it turns out that it is more efficient
to calculate elastic constants $C_{ij}$  by exploiting the 
relations~\cite{parrinello82} between
$C_{ij}$ and the thermal fluctuations of the strain tensor.

\begin{figure}[hbtp]
\begin{center}
\leavevmode
\hbox{ \epsfxsize=70mm \epsfbox{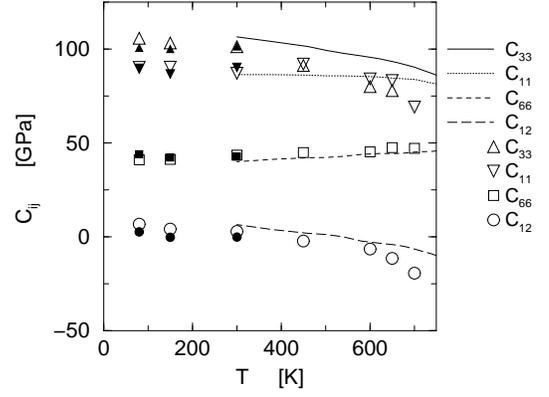} }
\begin{minipage}{8.5cm}
\caption{Various elastic constants. Experimental data is taken from
Carpenter et al. (1998). Open symbols refer to classical simulations,
filled symbols to PIMD simulations. Statistical error bars are
about 2~GPa.
\label{fig:ela_aq}
}
\end{minipage}
\end{center}
\end{figure}

Experimental, classical, and quantum mechanical
elastic constants are compared in Fig.~\ref{fig:ela_aq}.
Elastic constants can be expected to
show larger (relative) quantum corrections than lattice constants
and heat of formation~\cite{schoffel01}. For quartz, the reduction of
about 5 GPa in $C_{33}$ seems to be the most dramatic effect. 
At $T = 300$~K, classical and quantum mechanical elastic constants agree within
the statistical error bars. Below 300~K, the classical $C_{33}$ shows a 
stronger temperature dependence than the quantum  mechanical $C_{33}$.
This effect should be taken into account when trying to optimize
potential energy surfaces: $C_{33}$  predicted by the force field
parameters for $T = 0$~K should be a little larger than $C_{33}$
measured at a temperature of 300~K. For other $C_{ij}$ the same comment
applies in principle, but quantitatively, the effects are less dramatic.

\subsection{Dynamical Properties}
\label{subsec:dyn_prop}

As a generic dynamical property  we consider the (classical)
inverse-mass weighted momentum autocorrelation function $C(t)$ 
\begin{equation}
C(t) = \sum_i m_i^{-1} \left\langle \vec{p}_{i}(t) 
\vec{p}_{i}(0) \right\rangle,
\label{eq:time_cor}
\end{equation}
where $t$ denotes the real time.
$C(t)$'s Fourier transform $\tilde{C}(\omega)$ can be used to define
an effective density of states $g_{\rm eff}(\nu)$
\begin{equation}
g_{\rm eff}(\nu) = { \tilde{C}(2\pi\nu) \over Nk_BTn(\beta h\nu)}
\label{eq:dos}
\end{equation}
with $n(\beta h\nu)$ being introduced in Eq.~(\ref{eq:norm}).
$g_{\rm eff}(\nu)$ is identical with the real density of states (DOS)
if the harmonic approximation is valid.
$\tilde{C}(\omega)$ and hence
the effective DOS can be exactly related to the imaginary-time
correlation function $G(\tau)$
\begin{equation}
G(\tau) = \sum_i m_i \left\langle \left(\vec{R}_i(\tau) - \vec{R}_i(0)
\right)^2\right\rangle
\end{equation}
via the two-sided Laplace transform
\begin{eqnarray}
G(\tau) & = & \int_{-\infty}^{\infty} d\omega\,
\exp\left({-\hbar \omega \beta / 2}\right)
\nonumber\\ & &\, \times
 {\tilde{C}(\omega)\over \omega^2}
 \left[ \cosh\left\{\hbar\omega\left({\beta\over 2}-\tau\right)\right\} - 
\cosh{\hbar\omega\beta\over 2}\right].
\label{eq:twosided}
\end{eqnarray}
Note that the imaginary time $\tau$ has to be considered within the
interval $0 \le \tau < \beta$. Outside of this interval, imaginary-time
correlation functions are repeated periodically.

\begin{figure}[hbtp]
\begin{center}
\leavevmode
\hbox{ \epsfxsize=70mm \epsfbox{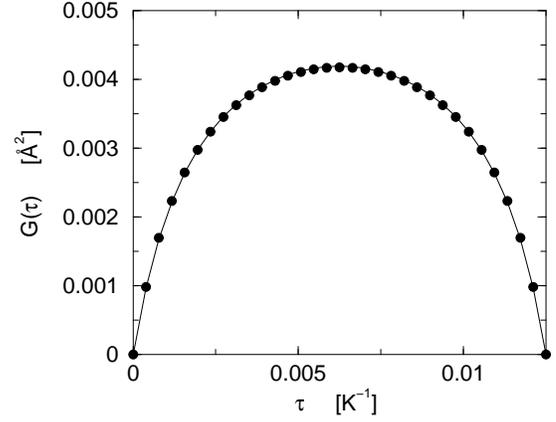} }
\begin{minipage}{8.5cm}
\caption{The imaginary-time correlation function $G(\tau)$ at $T = 80$~K
as obtained directly (points) and via the centroid PIMD
method.
\label{fig:imacorr}
}
\end{minipage}
\end{center}
\end{figure}

Eq.~(\ref{eq:twosided}) is useful to check the validity of the
centroid molecular dynamics (CMD) method and hence to establish
the validity of spectral functions as obtained by CMD.
If $C(t)$ is determined in terms of the mass-weighted autocorrelation function
of the centroid velocities and use is made of Eq.~(\ref{eq:centroid_spec}),
the effective DOS and hence $G(\tau)$ can be estimated in terms of centroid 
dynamics. If CMD is applicable, $G(\tau)$ as obtained by direct sampling
and $G(\tau)$ as estimated via CMD have to agree.
As shown in Fig.~\ref{fig:imacorr}, the agreement is perfect within
our statistical error bars for $\alpha$-quartz at very low temperatures.
Of course, this agreement could be expected as the dynamics are dominated
by the harmonic interactions unlike the thermal expansion coefficients.
Note that a  purely classical simulation leads to a similarly good agreement.
CMD and classical velocity autocorrelation functions can
barely be distinguished in the case of quartz.

\begin{figure}[hbtp]
\begin{center}
\leavevmode
\hbox{ \epsfxsize=80mm \epsfbox{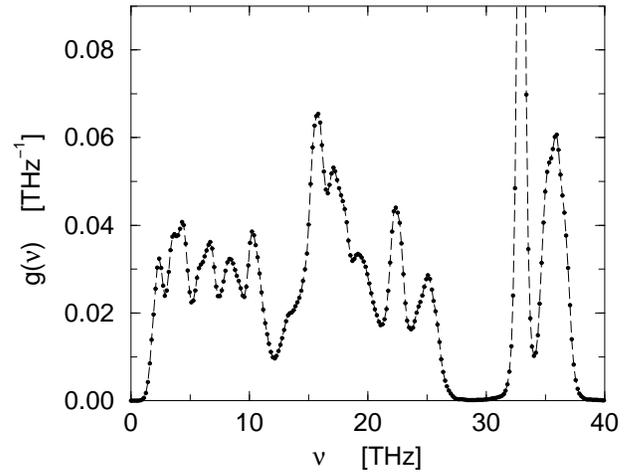} }
\begin{minipage}{8.5cm}
\caption{The density of states $g(\nu)$ as predicted by the centroid PIMD
method. $g(\nu)$ serves as an input spectrum to previos figure.
Dotted line is drawn to guide the eye.
\label{fig:dos}
}
\end{minipage}
\end{center}
\end{figure}

Fig.~\ref{fig:dos} shows the density of states as calculated via the
centroid PIMD. Of course, a degree of (meaningful) complexity 
in a spectrum such as shown in Fig.~\ref{fig:dos}
can never be obtained by inverting imaginary-time correlation functions
as shown in Fig.~\ref{fig:imacorr}. Thus, centroid PIMD are a useful tool
to obtain DOS of silica.

\section{Conclusions}
This study shows that path integral molecular dynamics (PIMD) are an efficient
tool to calculate low-temperature properties of solids even if the complexity
is larger than in rare gas crystals or other monoatomic solids. 
PIMD turns out to be particulary useful (as compared to path-integral 
Monte Carlo) when long-range forces have to be evaluated such as it is the
case for the simulations of silica.
Structural properties can be evaluated with high resolution and the shift
from properties that are obtained if atomic motion is treated classically
to the ``real'' quantum mechanical properties can be assessed.
This shift can also be calculated for elastic constants, which are notoriously
hard to compute even in classical simulations.

The result of the PIMD simulations anticipate that path integral techniques 
may not only become an important way of evaluating the merits and failures of
potential energy surfaces, but PIMD might give valuable input to 
{\it construct} reliable model potentials. Here, the PIMD calculations
of the thermomechanical properties of $\alpha$-quartz were based 
on the BKS potential~\cite{beest90}. The construction of the BKS potential
was pioneering in the sense that ab-initio calculations were combined with
bulk properties in order to fit the free model parameters. In the latter part,
lattice constants and elastic constants were calculated for a classical
system at $T = 0$~K from the (fit) parameters and adjusted such that agreement
with experimental ``quantum mechanical'' (finite temperatures)
data was optimum. The PIMD results in combination with the classical MD results
presented  in this paper, show that this part of adjusting the parameters
of the BKS potential allows for further optimization.
Of course, one can not necessarily expect to find a two-body potential energy
surface for silica that describes interactions much better than
the BKS potential. 


\label{sec:conclusions}

\acknowledgments
We thank Kurt Binder for useful discussions.
Support from the BMBF through Grant 03N6015 and
from the Materialwissenschaftliche Forschungszentrum  
Rheinland-Pfalz is gratefully acknowledged.

\newpage

\newpage

\end{multicols}
\end{document}